\documentclass[twocolumn,showpacs,preprintnumbers,amsmath,amssymb]{revtex4}

\usepackage{graphicx}
\usepackage{dcolumn}
\usepackage{bm}
\newcommand{\unity}{\ensuremath{{\rm 1 \negthickspace l}{}}}

\begin{document}


\title{Optimal Control-Based Efficient Synthesis of Building Blocks
of Quantum Algorithms\\[1mm] Seen in Perspective from Network Complexity towards Time Complexity\footnote{work presented in part at the QCMC 2004 in Glasgow}}

\author{T.~Schulte-Herbr{\"u}ggen}\email{tosh@ch.tum.de}
\author{A.~Sp{\"o}rl}
\affiliation{Department of Chemistry, Technical University Munich, Lichtenbergstrasse 4, D-85747 Garching, Germany}

\author{N.~Khaneja}
  \affiliation{Division of Applied Sciences, Harvard University, Cambridge MA02138, USA}

\author{S.J.~Glaser}
\affiliation{Department of Chemistry, Technical University Munich, Lichtenbergstrasse 4, D-85747 Garching, Germany\phantom{.}}

\date{\today}


\begin{abstract}
In this paper, we demonstrate that optimal control algorithms can be used to speed up the
implementation of modules of quantum algorithms or quantum simulations in
networks of coupled qubits. The gain is most prominent in realistic cases, where 
the qubits are not all mutually coupled. Thus
the shortest times obtained depend on the coupling topology as well as on the characteristic
ratio of the time scales for local controls {\em vs}
non-local ({\em i.e.}~coupling) evolutions in the specific experimental setting.
Relating these minimal times to the number of qubits gives the tightest
known upper bounds to the actual time complexity of the quantum modules.
As will be shown, time complexity is a more realistic measure of the experimental
cost than the usual gate complexity.

In the limit of fast local controls (as {\em e.g.} in NMR), time-optimised realisations
are shown for the quantum Fourier transform (QFT) and the multiply controlled {\sc not}-gate
({\sc c$^{n-1}$not}) in various coupling topologies of $n$ qubits. The speed-ups
are substantial: in  a chain of six qubits
the quantum Fourier transform so far obtained by optimal control
is more than eight times faster than the standard decomposition into controlled phase,
Hadamard and {\sc swap} gates, while the {\sc c$^{n-1}$not}-gate for completely coupled network of six
qubits is nearly seven times faster.
\end{abstract}

\maketitle


\section{Introduction}
A key motivation for using experimentally controllable quantum systems to
perform computational tasks or to simulate the behaviour of other quantum systems
\cite{Fey82, Fey96} roots in reducing the complexity of the problem when going
from a classical setting to a quantum setting. The most prominent
pioneering example being Shor's quantum algorithm of prime factorisation \cite{Shor94, Shor97}.
While classical prime factorisation algorithms are of non-polynomial complexity $NP$ \cite{Pap95},
Shor's quantum algorithm brings it down into the class of polynomial
complexity $P$. 
Another celebrated example is Grover's quantum search algorithm \cite{Grov96, Grov97},
which allows for searching in an unstructured data base of $n$ qubits with $N = 2^n$ items
in $O(\sqrt{N})$ quantum steps instead of $O(N)$ classical ones.

As a matter of fact, many quantum algorithms can be subsumised as solving 
{\em hidden subgroup problems} in an efficient way \cite{EHK04}.
In the abelian case, the speed-up hinges on the quantum Fourier transform (QFT):
while the network complexity of the fast Fourier transform (FFT) for $n$ classical bits
is of the order $O(n 2^n)$ \cite{CT65, Beth84}, the QFT for $n$ qubits shows a complexity
of order $O(n^2)$.
For implementing a quantum algorithm or a quantum simulation in
an experimental setup, it is customary to break it into 
universal elementary quantum gates \cite{Deu85}. Common sets comprise {\em e.g.} 
(i) local
operations such as the Hadamard gate, the phase gate and (ii) the
entangling operations {\sc cnot}, controlled-phase gates,
$\sqrt{\text{\sc swap}}$, $i$\,{\sc swap} as well as (iii) the {\sc swap} operation.
The number of elementary gates required for implementing 
a quantum module then gives the network or gate complexity.

However, gate complexity often translates into too coarse an estimate for the actual
time required to implement a quantum module (see {\em e.g.} \cite{VHC02, CHN03, ZGB}), 
in particular, if the time scales of a specific experimental setting have to be matched. 
Instead, effort has been taken to give upper bounds on the actual {\em time complexity} \cite{WJB02},
which is a demanding goal from the algebraic point of view.
With the time required for implementing a module in a 
specific experimental setting as the most realistic measure of cost,
here we use methods of optimal control theory to find the minimum time by
trying to solve the time-optimal control problem. The solution is
hard to come by in general, so here we resort to numerical algorithms.
The shortest times obtained depend on the coupling topology as well as on the characteristic
ratio of the time scales for local controls {\em vs}
non-local ({\em i.e.}~coupling) evolutions and thus embrace the specific experimental setting.
Relating these minimal times to the number of qubits gives the tightest
known upper bounds to the actual time complexity of the quantum modules
in a realistic experimental setup.

Moreover, as will be discussed, in the
generic case there is no simple one-to-one relation between time complexity
and network complexity, because of different time-scales between local
and non-local controls, different coupling topologies allowing for different
degrees of parallelisation, and different types of coupling interactions
matching different sets of elementary gates.

Thus here we leave the usual approach of decomposing gates into sets of 
discrete universal building blocks.  Instead, the scope is to exploit the differential geometry
of the unitary group for optimisation \cite{Science98,NMRJOGO} when
using the power of quantum control in order to
obtain constructive bounds to minimal time both as close to
the experimental setting and as tight as possible.
In the limit of zero cost for the fast local controls (as in NMR)
compared to the slow coupling interactions, we give
decompositions for the QFT and the multiply-controlled {\sc not}-gate 
{\sc c$^{n-1}$not} that are dramatically
faster than the fastest decompositions into standard gates known so far.

The paper is organised as follows: 
the first focus is on the fact that for time-optimal decompositions
of a desired unitary gate into a sequence of evolutions of experimentally
available controls the global phase may play a role. This is the case when
{\em e.g.} there are different time scales for local versus non-local
controls.  However, global phases can readily be absorbed by shifting
gradient flows from unitary to projective unitary groups.
Then numerical time-optimal control provides the currently best upper
bounds to the actual time complexity of quantum modules like the QFT
or the {\sc c$^{n-1}$not}-gate in various coupling topologies 
of $n$ qubit systems. Here we present examples with $n$ up to seven.
For $n\geq 3$, the resulting time complexities are 
bounded from above by $KAK$-type decompositions 
taken to sub-Riemannian regimes \cite{Khaneja01a, Khaneja01b, Khaneja02}.
Finally we give an outlook generalising the methods developed from
spin- to pseudo-spin systems.

Although the applications presented here refer to time-optimised quantum computing in
the NMR-limit of fast local controls, the methods introduced
are very general and apply to all systems whose dynamics can be cast into the closed
form of finite-dimensional Lie algebras (to sufficient approximation).

\section{Controllable Systems}
\subsection{Spin- and Pseudo-Spin Systems}
Here we address {\em fully controllable} \cite{JS72,SJ72,Bro72,Bro73,BW79,TOSH-Diss}
quantum systems represented as spin-
or pseudo-spin systems, {\em i.e.} those in which---neglecting 
decoherence---for any initial state represented by its density operator $A$,
the entire unitary orbit
$U(A):= \{UAU^{-1}\, |\, U \; {\rm unitary}\}$
can be reached \cite{AA03}.
In systems of $n$ qubits ({\em e.g.} spins-$\frac{1}{2}$), this is the case under
the following mild conditions \cite{TOSH-Diss,Science98,GA02}: 
(1) the qubits have to be inequivalent {\em i.e.} distinguishable and selectively addressable, and
(2) they have to be pairwise coupled ({\em e.g.} by Ising interactions), where the coupling topology may take
the form of any connected graph.

\subsection{Time Scales for Local and Non-Local Controls}
Let the quantum system evolve in a time interval $t_k$
under combinations of piece-wise constant 
control Hamiltonians $\{H_j\}$ and the drift $H_{\rm d}$, 
{\em i.e.}~the free-evolution Hamiltonian, according to
\begin{equation}\label{H_control}
H^{(k)} = H_{\rm d} + \sum_j u_j^{(k)} H_j^{(k)} \quad .
\end{equation}
In NMR spin dynamics \cite{EBW87}, for instance, 
the local controls on qubit $\ell$ are represented by a linear combination of
the Pauli matrices $\{\sigma_{\ell x}, \sigma_{\ell y}\}$.
And the drift term is governed by the weak scalar couplings
(reminiscent of Ising interactions)
\begin{equation}
H_{\rm d} = \pi \sum_{\ell<m} J_{\ell m}\; \tfrac{1}{2}\; \sigma_{\ell z}\otimes\sigma_{m z} \quad,
\end{equation}
provided the couplings between spins are much smaller than the difference between
the eigenfrequencies (shifts $\Omega$) of the respective spins:
$|J_{\ell m}| \ll |\Omega_\ell - \Omega_m|$.
This is the case in heteronuclear spin systems. And in quantum control even
the homonuclear ones can be designed such as to meet this greatly simplifying 
approximation \cite{MFM+00}.

For the system to be {\em fully controllable} in the sense
outlined above, $\{H_d\}\cup\{H_j\}$ has to form a generating set of the Lie 
algebra $su(N)$ by way of the Lie bracket.

Often the time scale for local controls is
considerably faster than for the costly slow coupling evolutions.

\section{Time-optimal Quantum Control}
In order to control a quantum system of $n$ qubits (spins-$\tfrac{1}{2}$) such as to
realise a quantum gate or module of some quantum algorithm given by the unitary propagator
$U_{\rm G} \in U(2^n)$ in minimal time, one has to decompose
\begin{equation}\label{U_seq}
U_{\rm G} \sim U(T) = 
e^{-it_M H^{(M)}} 
	\cdots e^{-it_k H^{(k)}} \cdots e^{-it_1 H^{(1)}}
\end{equation}
---up to a global phase factor---into a {\em time-optimal sequence} 
($T := \sum_k t_k \overset{!}{=} \min$) 
of 
evolutions under piece-wise constant Hamiltonians $H^{(k)}$.

\subsection{Relevance of Global Phases}
\begin{figure}[Ht!]
\includegraphics[scale=0.47]{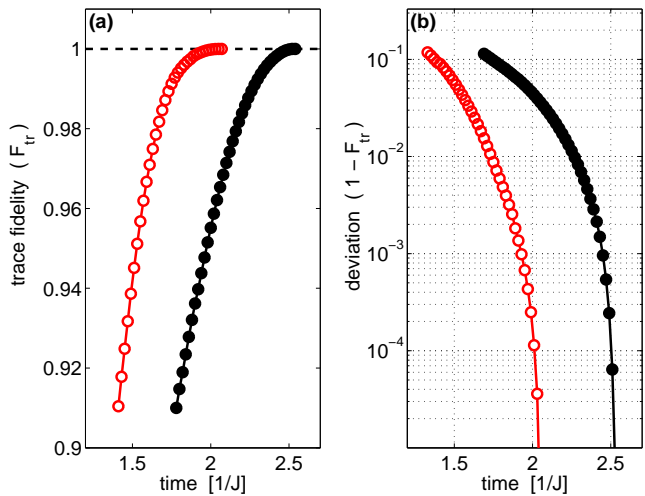}
\caption{\label{fig:phases} Global phase dependence of the
times needed to implement the 3-qubit QFT 
on a linear chain ($L_3$) of nearest-neighbour interactions with uniform 
weak scalar $J$-couplings. The right curves
($\bullet$) show the special unitary implementation of QFT with the smallest
global phase $\phi_0=\frac{1\pi}{16}$, where it takes $2.53\, J^{-1}$ to reach a trace 
fidelity $\geq 0.99999$. The left curves ($\circ$) display
the fastest QFT implementations obtained. They are attained with the global phase $\phi_1=\frac{5\pi}{16}$.
Trace fidelities $\geq 0.99999$ are reached after $2.05\, J^{-1}$. Times for local
controls are assumed to be negligible in this limit 
matching the typical NMR scenario, where the time cost is 
determined by coupling evolutions.
(a) gives the trace fidelities against time, while (b) shows the 
devitiations from full fidelity in a semilogarithmic way.}
\end{figure}
However, as propagators generated by the traceless
spin Hamiltonians are elements of 
the respective {\em special} unitary groups, the quantum gates $U_{\rm G}$ are realised by
$U(T)$ just up to global phases $\phi_p$
\begin{equation}
U_{\rm G} = e^{-i\phi_p} U(T) \quad;
\end{equation}
so $U_{\rm G} \in U(N)$, while $U(T) \in SU(N)$. 
For $n$ spins-$\tfrac{1}{2}$ read $N:=2^n$ henceforth. 
Note that with the centre of $SU(N)$ being
\begin{equation}
{\Bbb Z}_N := \{e^{i \frac{2\pi p}{N}}\;\unity_N\; |\; p = 0,1,\dots,N-1\}\quad ,
\end{equation}
one has a choice of $N$ such phases
\begin{equation}
\phi_p \in \{\phi_0 + \frac{2\pi p}{N} |\; p = 0,1,\dots,N-1\} \quad ,
\end{equation}
where $\phi_0$ shall be the smallest angle $\phi_0\in[0,\pi]$ so that
${\rm det}\{e^{i\phi_0} U_{\rm G}\} = +1$.
Although global phases are clearly immaterial to quantum evolutions
$\rho_0 \mapsto U\rho_0 U^{-1}$, it is important to note
they do in fact contribute substantially
to the over-all time needed to implement $ U_{\rm G}$:
consider, {\em e.g.},
\begin{equation}
e^{-i\tfrac{\pi}{2} \sigma_{\ell z}\otimes\sigma_{mz} }
	= e^{i\tfrac{\pi}{2} } e^{-i \tfrac{\pi}{2}(\sigma_{\ell z}\otimes\unity + \unity\otimes\sigma_{mz}) } 
		\quad,
\end{equation}
where the non-scalar part of the
right-hand-side can be realised solely by (fast) local controls,
whereas the left-hand-side hinges on nothing but (slow) coupling evolution.
 
In Fig.~\ref{fig:phases} this is further illustrated for the 3-qubit QFT
implemented on a chain of three spins connected by nearest-neighbour
interactions of weak scalar coupling in the NMR limit of zero time cost
for the fast local controls.

\subsection{Optimal Control on Projective Groups}
For a given unitary quantum gate $U_{\rm G}$ and propagators $U=U(t)$
describing the evolution of the quantum system, there are the two geometric
tasks, one that explicitly carries the phase, while the other one
automatically absorbs it as desired:
\begin{enumerate}
\item[(1)] {\em minimise the distance $\|U-U_{\rm G}\|_2^{\phantom{-}}$ 
	by maximising $\Phi_1 := {\rm Re\; tr} \{U^\dagger_{\rm G} U \}$};\\
\item[(2)] {\em minimise the angle $\measuredangle(U, U_{\rm G})$ {\rm mod}($\pi$)
	by maximising $\Phi_2 := | {\rm tr} \{U^\dagger_{\rm G} U \} |^2$}.
\end{enumerate}
(1) In terms of control theory, the first task is to maximise the quality functional
$\Phi_1[U(t)] = {\rm Re\; tr} \{U^\dagger_{\rm G} U(T) \}$
with $0\leq t \leq T$
subject to the equation of motion $\dot{U}(t) = -i H U(t)$ (with $H=H_d+\sum_j u_j H_j$)
and the initial condition $U(0)=\unity$, whereas the final condition $U(T)$ is
free at an appropriately fixed final time $T$ ({\em vide infra}).
As usual, the problem is readily solved by introducing the
operator-valued Lagrange multiplier $\lambda(t)$ satisfying
$\dot{\lambda}(t) = - iH \lambda(t)$
and a scalar-valued Hamiltonian function
\begin{equation}
h(U(t_k)) = {\rm Re\; tr} \big{\{}\lambda^\dagger(t_k) \big(-i(H_d+\sum_j u_j H_j)\big) U(t_k)\big{\}}\quad .
\end{equation}
Then, Pontryagin's maximum principle \cite{Pont64} may be exploited in 
a quantum setting \cite{Sam90,Kha+05} to require
\begin{equation}\label{control_law}
\frac{\partial h(U(t_k))}{\partial u_j} \equiv
-{\rm Im}\;{\rm tr}\{\lambda^\dagger(t_k) H_j U(t_k)\} \overset{!}{=} 0
\end{equation}
as well as the final condition for the adjoint system
\begin{equation}\label{lambda_T}
\lambda(T) = -\frac{\partial{\Phi_1(T)}}{\partial{U(T)}} = - U_{\rm G}
\end{equation}
thus allowing to implement a gradient-flow based recursion \cite{Kha+05}. For the
amplitude of the $j{\rm th}$ control in iteration $r+1$ at time interval $t_k$
one finds with $\alpha$ as a suitably chosen step size
\begin{equation}\label{controllaw}
u_j^{(r+1)}(t_k) = u_j^{(r)}(t_k) +
        \alpha \tfrac{\partial h(U(t_k))}{\partial u_j}\quad .
\end{equation}
The procedure is then repeated for a set of decreasing
final times $T$ up to a minimal  time $\tau$ still allowing to get sufficient fidelity
(compare Fig.~\ref{fig:phases}).

\vspace{3mm}
\noindent
(2) The second task amounts to maximising
$\Phi_2[U(t)] = | {\rm tr} \{U_{\rm G}^\dagger U(T)\} |^2$,
which is equivalent to the square of the trace fidelity and
is easy to handle by gradient-flow methods.
This problem, however, can readily be reduced to task (1): 
observe that to $U\in SU(N)$, 
\begin{equation}
\hat{U} := U^*\otimes U
\end{equation}
is a representation of
the corresponding element of the {\em projective special unitary group}
\begin{equation}
PSU(N) \overset{{\rm iso}}{=} \frac{SU(N)}{{\Bbb Z}_N} \overset{{\rm iso}}{=} \frac{U(N)}{U(1)}
\end{equation}
embedded in $SU(N^2)$.
Hence this representation is highly reducible yet very convenient, because
\begin{equation}
\Phi_1[\hat{U}(t)] = {\rm Re\; tr} \{\hat{U}_{\rm G}^\dagger \hat{U}(T)\}
		= | {\rm tr} \{U_{\rm G}^\dagger U(T)\} |^2 
		= \Phi_2[U(t)]\,.
\end{equation}
Hence one may adopt the previous results to obtain the
gradient flow on $PSU(N)$ just by using
\begin{equation}
\frac{\partial h(\hat{U}(t_k))}{\partial u_j} \equiv -2\, {\rm Im}\; \big({\rm tr}\{\lambda^t(t_k) H^t_j U^*(t_k)\} 
	\cdot {\rm tr}\{\lambda(t_k)^\dagger U(t_k)\}\big)
\end{equation}
in Eqn.~\ref{controllaw}.
Thus an explicit tensor product never enters the algorithm.
And the final condition of the adjoint system does not
require any prior knowledge or screening of the global phase 
ultimately giving the fastest implementation as has been the case
in previous settings, {\em e.g.} \cite{Wes04},
because embedding $PSU(N)$ in $SU(N^2)$ enforces a global phase of zero.
Absorbing the phases cuts the number of computations 
for $n$-qubit systems by a factor $N=2^n$.

Having reduced task (2) to task (1) also saves all the convergence and step-size
considerations \cite{NMRJOGO} from $SU(N^2)$ to apply to $PSU(N)$.

With these stipulations, the Hamiltonians $H_k$ according to Eqn.~\ref{H_control},
and the numbering as in Eqn.~\ref{U_seq},
the iterations $r$ of Eqn.~\ref{controllaw}
can be used in the following algorithmic scheme \cite{Kha+05}:
\begin{enumerate}
\item set initial controls $u_j^{(0)}(t_k)$ for all times $t_k$ with \\
	$k=1,2,\dots M$ at random or by guess;
\item starting from $U_0=\unity$, calculate for all $t_1,t_2,\dots t_k$
the forward-propagation
\begin{equation}
\begin{split}
\phantom{X}\quad
U^{(r)}(t_k)\; =\; &e^{-i(t_k-t_{k-1}) H_k^{(r)}} e^{-i(t_{k-1}-t_{k-2}) H_{k-1}^{(r)}}\; \dots \\
	&\quad\times\;e^{-i(t_1-t_0) H_1^{(r)}} U_0
\end{split}
\end{equation}
\item likewise, starting with $T=t_M$ and $\lambda(T)$ from Eqn.~\ref{lambda_T},
	compute for all $t_M, t_{M-1}, \dots t_k$
	the back-propagation
\begin{equation}
\begin{split}
\phantom{X}\quad
\lambda^{(r)}(t_k)\; =\; &e^{i(t_k-t_{k-1}) H_k^{(r)}} e^{i(t_{k+1}-t_k) H_{k+1}^{(r)}}\;\dots \\
	&\quad\times\; e^{i(t_M-t_{M-1}) H_M^{(r)}} \lambda(T)
\end{split}
\end{equation}
\item calculate $\frac{\partial h(U(t_k))}{\partial u_j}$ according to Eqn.~\ref{control_law};
\item with $u_j^{(r+1)}(t_k)$ from Eqn.~\ref{controllaw} update all the piece-wise
Hamiltonians to $H_k^{(r+1)}$ and return to step $2$.
\end{enumerate}

\section{Applications}
\begin{figure}[Ht!]\label{fig:topologies4}
\includegraphics[scale=0.43]{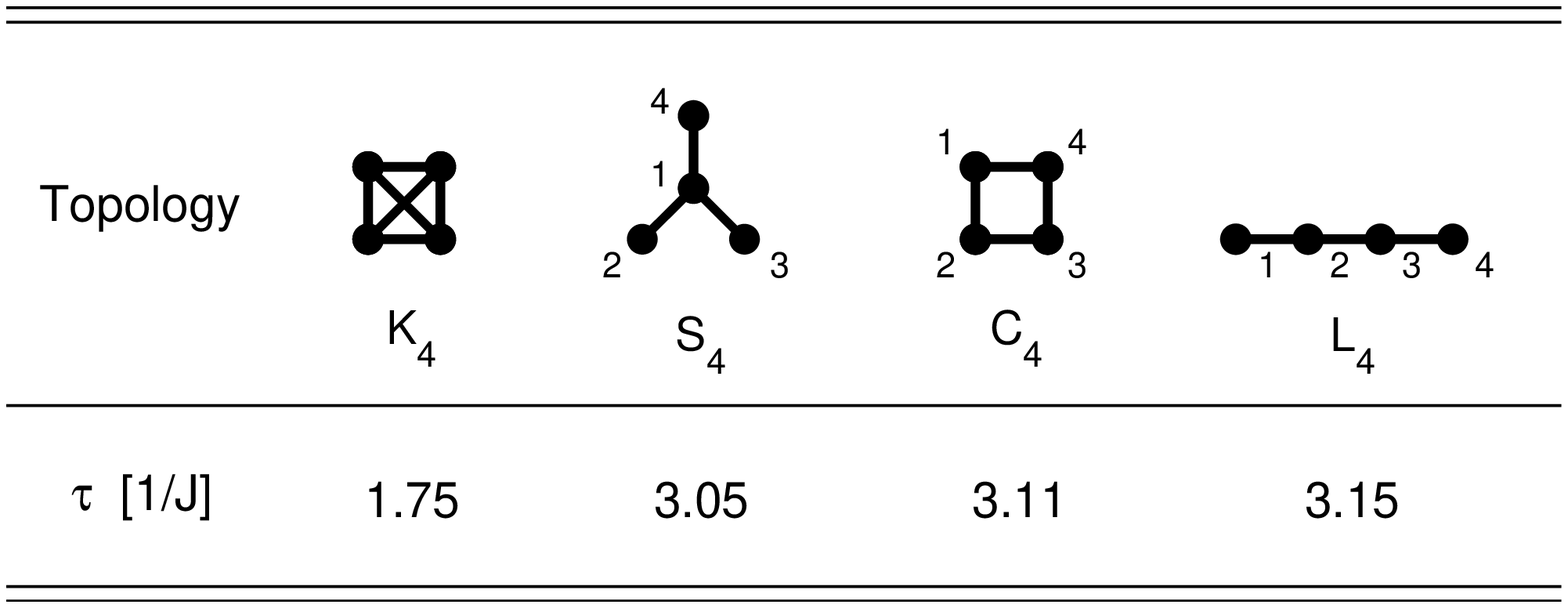}
\caption{\label{fig:topo_table} Four ordered connected graphs with 4 vertices
representing the topology of pairwise couplings (edges) between 4 qubits (vertices).
Times given 
are for the shortest QFT-realisations obtained by numerical 
time-optimal control rounded to $0.01\, J^{-1}$.}
\end{figure}
For simplicity, the coupling strengths in all the subsequent examples
are assumed uniform, thus enabling to give the times in units of
$J^{-1}$. However, all our algorithms can equally well cope with
non-uniform coupling constants directly matching the experimental
settings.

\subsection{Towards a Time-Optimal Quantum Fourier Transform}
The quantum Fourier transform (QFT) is central to all quantum algorithms of
abelian hidden subgroup type \cite{NC00, ABH+01}. 
The time required to implement this module in $n$-qubit systems clearly
depends on the topology of the coupling interactions. 

Fig.~\ref{fig:topo_table} shows some of the topologies for the couplings of
four qubits and the respective
times (best numerical results from optimal control)
for implementing the 4-qubit QFT.
Clearly, the complete coupling topology corresponds to the
maximal graph $K_n$ and thus allows for the fastest implementation,
while the chain of nearest-neighbour interactions $L_n$ is the
minimal connected graph entailing the slowest implementation.
Note, however, that the minimal times also depend on the ordering
in the graph, because permutations (carried out by transpositions) 
may call for timewise costly {\sc swap}s.

The decomposition into standard gates (controlled phase gate,
Hadamard, and {\sc swap}) can readily be made time-optimal 
only in complete coupling topologies ($K_n$).
There the minimal time can easily be expressed
in units of $J^{-1}$ as a function of the number qubits
(compare \cite{PHC00,SuterFT})
\begin{equation}
\tau(n) = \frac{1}{4} (n + 3)\quad,
\end{equation}
where the constant covers the final {\sc swap}.

\begin{figure}[Ht!]
\includegraphics[scale=0.43]{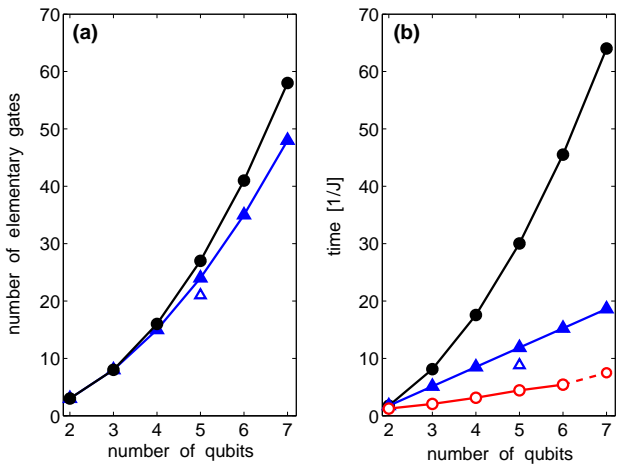}
\caption{\label{fig:qftln} (a) Gate complexity of the QFT in linear coupling topologies $L_n$. 
Standard-gate decomposition ($\bullet$) \cite{Saito} and optimised scalable
gate decomposition ({\protect\small{$\blacktriangle$}}) \cite{Blais}.
(b) Time complexity of the QFT in linear coupling topologies.
Upper traces give analytical times associated with the decompositions of part (a):
standard-gate decompositions ($\bullet$) \cite{Saito}
and optimised scalable gate decompositions ({\protect\small{$\blacktriangle$}}) \cite{Blais};
({\protect\small{$\vartriangle$}}) gives a special ({\em i.e.}~non scalable) 
five-qubit decomposition into standard gates obtained by simulated annealing \cite{Blais}.  
Lowest trace: speed-up by time-optimal control with shortest numerical realisations 
obtained ($\circ$) rounded to $0.01\, J^{-1}.$
Further details in Tab.~\ref{tab:qft_table}. 
(At $7$ qubits current decompositions
show a trace fidelity of $\sim 0.99$ and thus have not enough significant digits
to be included in the subsequent table.)}
\end{figure}
\begin{table}
\caption{\label{tab:qft_table}
Speed-up of the Quantum Fourier Transform on Linear Spin Chains, $L_n$}
\begin{ruledtabular}
\begin{tabular}{cccccc}
qubits 
& stand. QFT\footnote{analytical times for decomposition into standard gates
\cite{Saito}} 
& Blais\footnote{\cite{Blais} in brackets: the non-scalable special 5-qubit QFT}
&best results\footnote{upper bounds to minimal time for achieving a trace
                fidelity of $\geq 0.99999$ by numerical optimal control} 
& \multicolumn{2}{c}{speed-ups} \\
& $\tau \;[1/J]$\footnote{times $\tau$ are rounded to $0.01\, J^{-1}$}
& $\tau \;[1/J]^d$
& $\tau \;[1/J]^d$
& stand. & Blais\\[1mm]
\hline\\[-2.5mm]
2 & 1.75 & 1.75  & 1.25  &1.40 &1.40\\
3 & 8.13 & 5.13  & 2.05  &3.94 &2.50\\
4 &17.56 & 8.50  & 3.15  &5.58 &2.70\\
5 &30.03 & 11.88(8.81) & 4.44  &6.77 &2.67(1.98)\\
6 &45.52 & 15.25 & 5.43  &8.38 &2.81\\
\end{tabular}
\end{ruledtabular}
\end{table}
However, as is shown in
Fig.~\ref{fig:qftln}
and Tab.~\ref{tab:qft_table} (note the details in the table caption),
in linear spin chains ($L_n$)
with nearest-neighbour Ising interactions, time-optimal control
provides a decomposition of the QFT that is much faster than the corresponding
decomposition into standard gates would impose: in six
qubits, for instance, the speed-up is more than eight-fold and
in seven qubits approximately nine-fold.

For a fair comparison, however, note that Blais \cite{Blais} permutes 
output qubits for saving {\sc swaps}. 
However, searching through $n!$ permutations is beyond our cpu-time credits,
but may provide even faster realisations in the future.

\subsection{Towards a Time-Optimal {\sc c$^{n-1}$not}}
\begin{figure}[Ht!]
\includegraphics[scale=0.43]{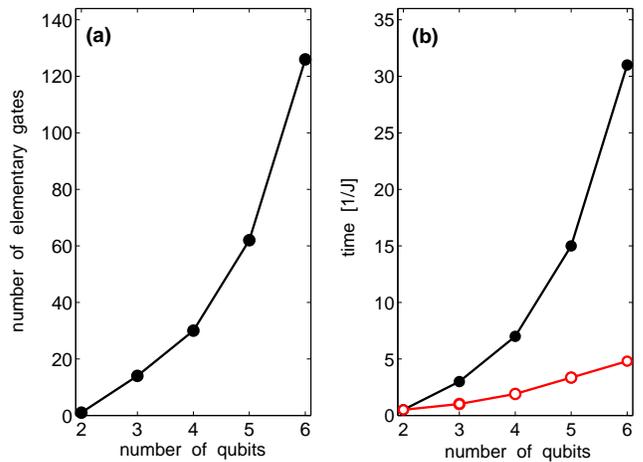}
\caption{\label{fig:Cn_NOT} 
(a) Network complexity of the {\sc c$^{n-1}$not}-gate on complete coupling topologies $K_n$ \cite{Barenco}. 
(b) Time complexity of the {\sc c$^{n-1}$not}-gate on complete coupling topologies.
Upper trace: analytical times for decomposition into standard
gates ($\bullet$) \cite{Barenco}. Lower trace: speed-up by time-optimal control
with shortest times ($\circ$) currently needed for realising
{\sc c$^{n-1}$not} by numerical control rounded to $0.01\, J^{-1}$.}
\end{figure}

Likewise, one may strive to implement the {\sc c$^{n-1}$not}-gate in a time-optimal
way. In a complete coupling topology of $n$ qubits, the algorithmic complexity was described
by Barenco {\em et al.} \cite{Barenco} as increasing exponentially up to six qubits,
whereas the increase from seven qubits onwards was said to be quadratic.
Again, time-optimal control provides a dramatic speed-up in this case as well,
see Fig.~\ref{fig:Cn_NOT} and Tab.~\ref{tab:Cn_NOT_table} as well as the controls
in Fig.~\ref{fig:all-controls}. 

\begin{table}
\caption{\label{tab:Cn_NOT_table}Speed-up for the {\sc c$^{n-1}$not}-Gate 
in Complete Coupling Topologies of $n$ Qubits, $K_n$}
\begin{ruledtabular}
\begin{tabular}{cccc}
qubits
&stand. decomposition\footnote{Barenco {\em et al.} \cite{Barenco}}
&best results\footnote{upper bounds to minimal time for trace fidelities $\geq 0.99999$ 
	(for 6 qubits currently: $\geq 0.999$) by numerical optimal control} 
& speed-up\\
& $\tau \;[1/J]$\footnote{times $\tau$ are rounded to $0.01\, J^{-1}$}
& $\tau \;[1/J]^c$
&\\[1mm]
\hline\\[-2.5mm]
2 &0.5   &0.50	&1.00\\
3 &3.0   &1.01  &2.97\\
4 &7.0   &1.90 	&3.68\\
5 &15.0  &3.37	&4.45\\
6 &31.0  &(4.59)&(6.75)\\
\end{tabular}
\end{ruledtabular}
\end{table}

\subsection{Beyond Spins: Controlling Coupled Charge Qubits in Josephson Devices}
Obviously the optimal control methods presented thus far can be generalised such as to
hold for systems with finite degrees of freedom allowing for a pseudospin
formulation in terms of closed Lie algebras. Suffice it to mention
the standard {\sc cnot}-gate can be realised in
two coupled charge qubits of a solid-state Josephson device
some five times faster than in the pioneering setting of Nakamura \cite{Nak03}.
Yet one easily obtains a trace fidelity beyond $0.99999$ as will be shown elsewhere. 
With the same fidelities one finds realisations
of the {\sc Toffoli}-gate in three linearly coupled charge qubits
that are some nine times faster than by standard gate decomposition.

\section{Discussion}
\begin{figure}[Ht!]
\includegraphics[scale=0.47]{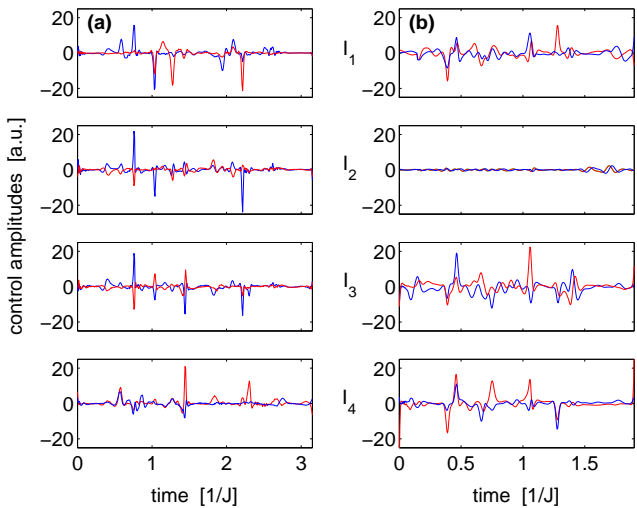}
\caption{\label{fig:all-controls}
Time course of controls for the shortest realisations obtained for
the following quantum modules: (a) the QFT on a linear
coupled four-qubit system ($L_4$);
(b) the {\sc c$^{3}$not}-gate on a fully coupled four-qubit system $K_4$.
Traces in blue: amplitudes for $\sigma_{\ell x}$-controls ($x$-pulses); 
red: amplitudes for $\sigma_{\ell y}$-controls ($y$-pulses) on the spins $\ell = 1,2,3,4$.}
\end{figure}
The goal is to extend the optimal control methods to larger 
modules of quantum algorithms or
simulations in order to implement them both
in a time-optimal and experimentally robust way.
Thus the growing set of numerical examples will hopefully
provide inspiration to understand time-optimal steerings of
quantum systems algebraically, which, however, seems very demanding in the
cases presented here (compare Fig.~\ref{fig:all-controls}).
In other instances such as for the propagator
\begin{equation}
U(t) = e^{-i \pi J t \;(\tfrac{1}{2}\sigma_{1z}\otimes\sigma_{2z}\otimes\sigma_{3z})}\quad,
\end{equation}
the theory is fully understood, and the
predictions based on sub-Riemannian
geodesics \cite{Khaneja02} perfectly match ({\em i}) the time-complexity
as well as ({\em ii}) the actual time course for the controls \cite{Kha+05}
for all $\pi J t \in [0,\tfrac{\pi}{2}]$.

Along these lines, the above controls may finally trigger
a theoretical understanding.
The ultimate challenge then is to extract a principle for
a {\em scalable} control scheme from the set of numerical
examples.

\section{Conclusion}
Here we have left the usual approach of decomposing quantum modules into 
sets of discrete building blocks, such as elementary universal quantum gates thus expressing
the cost as algorithmic network complexity.
Instead we proposed to refer to {\em time complexity} as the experimentally relevant cost:
it allows for exploiting the continuous differential geometry
of the unitary Lie-groups as well as the power of quantum control for getting
constructive upper bounds to the time complexity both perfectly matching
the experimental setting while being as tight as possible,
in particular when local and non-local operations are of
different time scale.
In the limit of zero cost for the fast local controls
we gave decompositions for the QFT and the multiply-controlled {\sc not}-gate
{\sc c$^{n-1}$not} that are dramatically
faster than the best decompositions into standard gates known so far
would impose. However, there is no guarantee the ultimate time optimum
is attained, also because permutations of the qubits may give further improvement.

The approach also clearly shows that in the
generic case there is no simple one-to-one relation between time complexity 
and network complexity. This is for very practical reasons: typically
(1) not all the elementary gates are of the same time cost, but each
experimental implementation comes with its characteristic ratio of
times required for local {\em vs} non-local (coupling) operations;
(2) not all the elementary gates have to be performed sequentially,
but can be rearranged so that some of the commuting operations
({\em e.g.} controlled phase gates between several qubits) or 
operations in disjoint subspaces can be parallelised;
(3) the coupling topology between the qubits does not have
to form a complete graph ($K_n$) but may be just a connected
subgraph,
and each graph comes with a specific potential of parallelising timewise
costly interactions; this is demonstrated for the QFT on complete coupling
topologies $K_n$ versus the linear coupling topology $L_n$, where
the parallel performance of controlled phase gates \cite{Blais} reduces
quadratic time complexity to linear complexity, which, however, can
be further speeded up by time-optimal control;
(4) the experimental setting with its specific type and individual strengths of coupling
interaction ({\em e.g.} Ising or Heisenberg-$XY$ or $XYZ$ type) related to the choice of
universal gates for the network decomposition may introduce some arbitrariness.

It is for these very reasons that time complexity is the more realistic
measure of the experimentally relevant cost than network complexity is.

\section{Outlook}
Although extrapolation may be premature, it is fair 
to anticipate that in systems of 
some 20 qubits network decompositions will often become
impractical. Thus time-optimal decompositions into 
controls actually available in the experimental setting
promise to widen the range of experimentally accessible
tasks significantly and will prove useful in many experimental 
implementations.  Moreover, analysing the topology-dependence of minimal times 
while allowing for non-uniform coupling strengths
will contribute valuable guidelines for designing optimised
architectures of quantum computational hardware.

By parallelising routines the results are 
currently being extended to more qubits 
so that time complexities can be deduced from fitting times
against number of qubits with confidence.

\begin{acknowledgments}
This work was supported in part by {\em Deutsche Forschungsgemeinschaft}, DFG,
in the grant Gl~203/4-2 and in the incentive SPP 1078 
({\em Schwerpunkt-Programm \/`Quanteninformations\-verabrbeitung\/', QIV}).
Encouraging discussion with Ivan Deutsch and Andrew Landahl at the QCMC 2004 is gratefully acknowledged.
\end{acknowledgments}
\bibliography{control21}
\end{document}